\documentclass[preprint,aps,prl]{revtex4}
\usepackage{graphicx}

\begin{document}

\title{Surface effects on the orbital order in the single layered manganite La$_{0.5}$Sr$_{1.5}$MnO$_4$}
\author{Y.~Wakabayashi$^{1,2}$, M.H.~Upton$^2$, S.~Grenier$^{2,3}$, J.P.~Hill$^2$, C.S.~Nelson$^4$, J.-W.~Kim$^{5}$, P.J.~Ryan$^{5}$, A.I.~Goldman$^{5}$, H.~Zheng$^{6}$, and J.F.~Mitchell$^{6}$}
\address{$^1$Photon Factory, Institute of Materials Structure Science, High Energy Accelerator Research Organization, Tsukuba 305-0801, Japan\\
$^2$CMPMS, Brookhaven National Laboratory, Upton 11973, NY, USA\\
$^3$Institut N\'eel, CNRS \& Universit\'e Joseph Fourier, BP 166, F-38042 Grenoble Cedex 9, France\\
$^4$National Synchrotron Light Source, Brookhaven National Laboratory,
Upton 11973, NY, USA\\
$^5$Ames Laboratory and Department of Physics and Astronomy, Iowa State Univ., Ames, 50011, IA, USA\\
$^6$Argonne National Laboratory, Argonne, 60439, IL, USA\\
}

\date{\today}

\begin{abstract}
We report the first observation of `orbital truncation rods' --- the scattering arising from the termination of bulk orbital order at the surface of a crystal. The x-ray measurements, performed on a cleaved, single-layered perovskite, La$_{0.5}$Sr$_{1.5}$MnO$_4$, reveal that while the crystallographic surface is atomically smooth, the orbital `surface' is much rougher, with an r.m.s. deviation from the average `surface' of $\sim$7\AA.\/ The temperature dependence of this scattering shows evidence of a surface-induced second order transition.
\end{abstract}

\maketitle 

An issue destined to be of increasing importance in the science of correlated-electrons systems in the coming years is the question of whether the electronic behaviour, such as charge, orbital or spin order, is different at a surface or interface compared to the bulk of the material. This question is of central importance from an applied perspective --- any device one fabricates from these materials will require electron transport across an interface; from a nanoscience perspective --- nanoscaled objects are defined by their very large surface-to-volume ratio; and from a fundamental perspective --- understanding the role of reduced dimensionality in determining the electronic behaviour of such systems is a key component in understanding the general problem of strongly correlated electron systems.

In many ways manganites represent an ideal system to address the role of the surface, because all the relevant degrees of freedom --- charge, spin and orbital --- play an active role in determining the ground state. As a result, they are exquisitely sensitive to perturbations, and hence, might be expected to exhibit relatively large surface effects. In fact, while the vast majority of experimental work has focused on bulk properties\cite{review1,review2}, the very few surface sensitive studies do suggest that the surface of a manganite may be quite different from the bulk. For example, the measurements of Freeland {\it et al.}\cite{Freeland05NatMat} showed that the top bilayer of a cleaved surface of a ferromagnetic bilayered manganite is magnetically dead, though the second bilayer has full bulk magnetic order. Similarly, the half-doped single-layer manganite, which is known to exhibit bulk orbital order below $T_{OO}=231$~K, showed no superlattice reflection corresponding to such ordering down to 80~K, when probed with surface-sensitive LEED \cite{Plummer01ProgSurfSci}.
Thus, there is evidence that the surface of manganite crystals may exhibit different electronic phases than the bulk --- a phenomena that has been coined `electronic reconstruction'\cite{Ohtomo02Nature,Okamoto04Nature} in analogy with more traditional surface science studies of atomic reconstruction driven by surface energetics. 

In this paper, we present the first study of orbital order at a surface, utilizing surface x-ray scattering. 
 The advantage of using x-ray scattering in this endeavour is that it provides information on both the electronic surface (specifically, atomic displacements associated with orbital order) and the chemical surface, separately. It is well known that the abrupt truncation of a sample by its surface gives rise to rod-shaped x-ray scattering around the Bragg reflections, the so-called crystal truncation rods (CTR). The intensity profiles of such rods provide detailed, quantitative information about the chemical surface\cite{Andrews85JPhysC}. Similarly then, the abrupt termination of orbital order by a surface should give rise to orbital truncation rods (OTR) running through the orbital superlattice peaks. These would allow one to quantify the surface orbital order. 
Here, we present the first observation of such
orbital truncation rod scattering. This allows us to quantitatively
probe the orbital order in the vicinity of the surface. Working with a
freshly cleaved La$_{0.5}$Sr$_{1.5}$MnO$_4$, we show that the orbital order has a significantly rougher interface than the chemical surface and a shorter
correlation length than in the bulk. These results imply that the
surface order parameter is reduced relative to the bulk. Further, the
temperature dependence of the surface orbital order suggests that the
surface exhibits a more continuous transition than the bulk ordering
transition. Taken together, these results demonstrate that the surface
is a significant perturbation in regard to the electronic phenomena of
orbital ordering and point to the need for an improved theoretical
understanding of how the surface couples to such electronic orders.

\begin{figure}
\includegraphics[width=10cm]{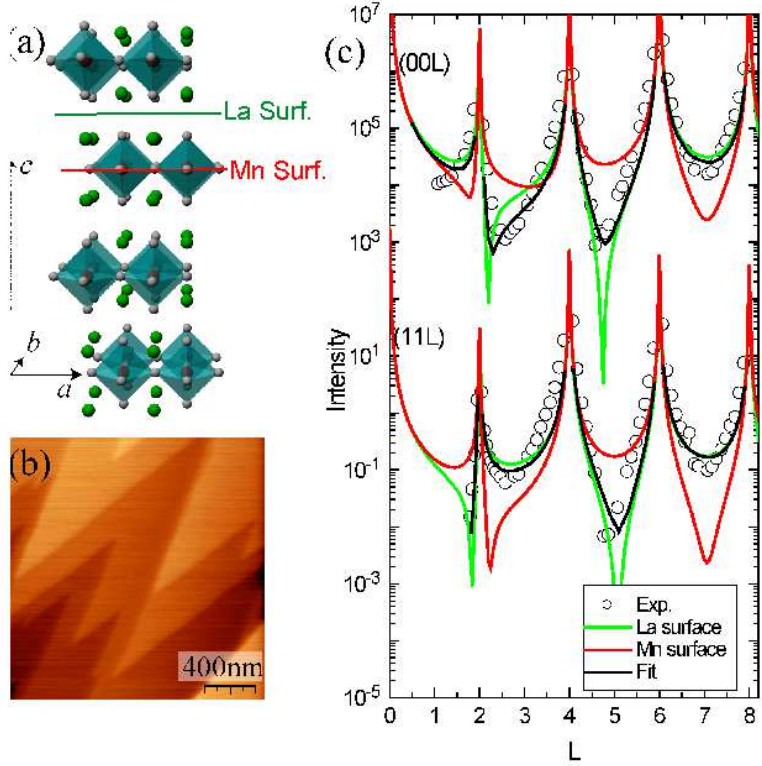}
\caption{(a) The crystal structure of La$_{0.5}$Sr$_{1.5}$MnO$_4$. Green, red and white spheres denote La/Sr, Mn, and O atoms, respectively. (b) AFM image of a cleaved sample used in this study. The step height was 6.2 $\pm$ 0.2 \AA. (c) (00$L$) and (11$L$) CTR scattering intensity distribution together with the calculated intensity for La and Mn terminated surfaces. The result of the curve fitting shows the chemical surface is predominantly La terminated.}
\label{fig:chem_sur}
\end{figure}

We first discuss surface preparation and characterization. Single crystal samples of La$_{0.5}$Sr$_{1.5}$MnO$_4$, which has the layered
perovskite structure shown in Fig.~\ref{fig:chem_sur} (a), were grown by
the floating zone method. The samples were then either cleaved perpendicular to the $c$-axis in He and placed in a sealed He-filled can, or cleaved in air, and put into a closed-cycle refrigerator. Both treatments gave the same results, indicating that the surface is stable to brief exposure to air, a result also found elsewhere\cite{Freeland06APS}. Figure~\ref{fig:chem_sur}(b) shows an atomic force microscope (AFM) picture of a freshly cleaved sample. The terraces seen here are on the order of 1 $\mu$m in size, much larger than the coherence length of the x-rays. The step height observed here was 6.2\AA $\pm$ 0.2 \AA, corresponding to $c/2$. We note that similar AFM pictures were obtained at the end of a week long x-ray experiment in vacuum. Analysis of the CTR intensity allows a determination of the precise termination of the cleaved sample\cite{You92PRB}. Figure \ref{fig:chem_sur} (c) shows the CTR scattering intensity distribution along the (00$L$) and (11$L$) directions together with the calculated intensity distribution for a La-terminated surface and a Mn-terminated surface (Fig.\ref{fig:chem_sur} (a)). The experimental data are well reproduced by assuming a predominantly La-terminated surface with very little surface relaxation.
 For comparison, there are conflicting reports in the literature for the termination of thin films, with both La and Mn termination reported\cite{Choi99PRB}. 

\vspace{1cm}

\begin{figure}
\includegraphics[width=7cm]{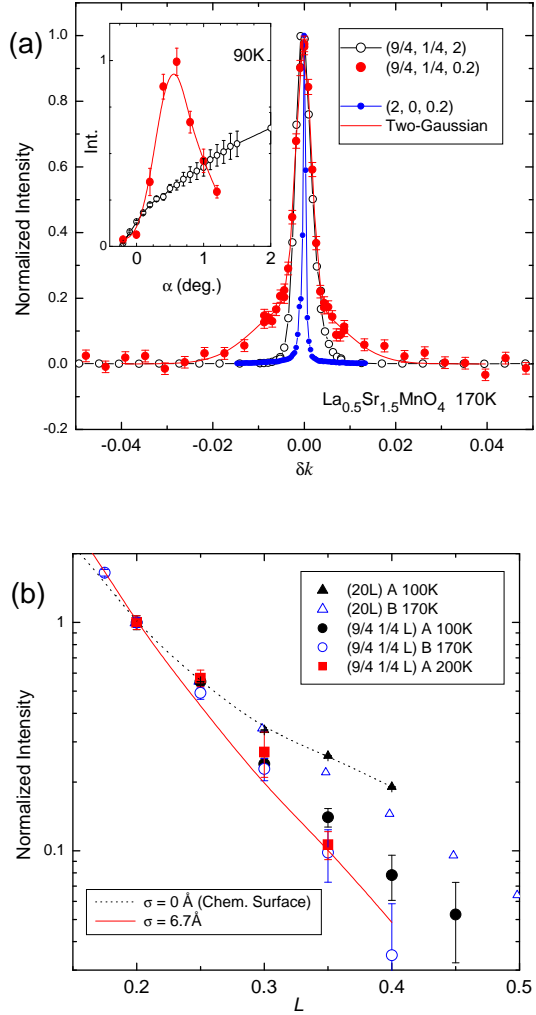}

\caption{(a) Transverse ($k$) scans through (2 0 0.2), ($\frac 94 \frac 14 2$), and
($\frac 94 \frac 14 0.2$) reciprocal lattice positions shown with blue circles, open circles, and red circles, respectively. The red curve shows the result of a two-Gaussian fitting for ($\frac 94 \frac 14 0.2$) profile. (inset) The incidence angle dependence of the scattered intensity at ($\frac 74 \frac {\bar{1}}{4} 2$) and ($\frac 74 \frac {\bar{1}}{4} 0.3$). The enhancement of the latter at $\alpha$ around $\theta_C=0.4^\circ$ shows that it results from surface scattering. (b) (20$L$) and ($\frac 94 \frac 14 L$) truncation rod intensity for samples A and B at 100K, 170K and 200K. The solid line shows the calculated intensity
distribution given by the measured I(20$L$) for sample A multiplied by $\exp(-\sigma^2q_z^2)$ with surface roughness parameter $\sigma=6.7$ \AA\/ (see text).}
\label{fig:OO-CTR}
\end{figure}

As discussed above, the structure of the chemical surface is reflected in the scattering along the ($hkL$) direction, where $h$, $k$ are integer multiples of reciprocal lattice vectors, that is rods along the surface normal $c$*-direction through the allowed, bulk Bragg reflections. Similarly, scattering originating from the surface of the orbital ordering must be present in `orbital truncation rods' --- rods of scattering along the same direction but through the superlattice reflections related to the orbital order, i.e. ($h+\frac14, k+\frac14, L$) where ($\frac14, \frac14, 0$) is the orbital ordering wave-vector\cite{Dhesi04PRL}.
One of the primary goals of this work was to observe such scattering. To this end, we measured the (20$L$)$\pm$($\frac 14 \frac 14 0$) intensity
distribution below $T_{OO}=$231~K, the bulk orbital order transition temperature.
We discuss first transverse scans taken across such `orbital rods' (figure~\ref{fig:OO-CTR}(a)).
For reference a transverse (2 $\delta k$ 0.2) scan profile across the truncation rod arising from the crystallographic surface is also shown (blue data). It is as sharp as the instrumental resolution. As discussed below, this is indicative of a smooth atomic surface. The open circles
show the profile of the ($\frac 94 \frac 14 2$) bulk orbital order Bragg peak. The finite width observed here shows that the correlation length of the bulk
orbital order is finite, and in fact, we find $\xi_{OO}^{\scriptsize{\rm bulk}}=$ 410\AA $\pm$ 8\AA, in agreement with earlier work\cite{Dhesi04PRL}. The transverse scan taken at ($\frac 94 \frac 14 0.2$), that is across the expected OTR is shown in red. A peak is clearly observed at this position. However, this peak is not, by itself, evidence of surface scattering. Such a peak could arise from stacking faults or other defects, in the bulk orbital order which would result in a finite correlation length for the orbital order along the $c$-direction and hence a tail along the $c$*-direction for superlattice reflections.

We can, however, distinguish surface scattering from such bulk scattering by
measuring the scattering intensity as a function of incidence angle $\alpha$. Surface scattered intensity will be enhanced at $\alpha = \theta_C$, the critical angle for total reflection, as a result of the interference between the incident and the scattered waves of the photon\cite{Feidenhansl89SurSciRep}. The inset to Fig.~\ref{fig:OO-CTR}(a) shows the scattered intensity as a function of $\alpha$ at the bulk orbital superlattice reflection ($\frac 74 \frac {\bar{1}}{4} 2$) and on the nominal OTR at ($\frac 74 \frac {\bar{1}}{4} 0.3$). The intensity at ($\frac 74 \frac {\bar{1}}{4} 0.3$) has a maximum around $\theta_C=0.4^\circ$ while a smooth increase is seen at ($\frac 74 \frac {\bar{1}}{4} 2$) with increasing $\alpha$, a result of the fact that more of the beam is intercepted by the sample at higher angles. Since the exit angle $\beta$ for ($\frac 74 \frac {\bar{1}}{4} 0.3$) is 3 to 1.5 degrees in this scan region, the observed intensity maximum cannot be due to a maximum in the sample illumination and detector acceptance, which happens at $\alpha = \beta$ in our experimental configuration.
Therefore, we conclude that the scattering observed at (2 0 0.2)$\pm$($\frac 14 \frac 14 0$), shown in Fig.~\ref{fig:OO-CTR} (a) does indeed come from surface
scattering and not from stacking faults or other bulk effects. This is the first such observation of the orbital truncation rod scattering.

Having established that such scattering originates from the truncation of the orbital order, we next discuss what can be learned from this new scattering. As shown in Fig.~\ref{fig:OO-CTR} (a), the transverse scan profile of
the rod intensity for the chemical surface and for the orbital surface are quite different. This difference in the scattering profiles directly reflects a difference in the surface roughness.\cite{Andrews85JPhysC} The sharp single peak observed for the chemical CTR scattering implies a very flat surface, and is consistent with the atomically smooth chemical surface observed in the AFM picture (Fig.~\ref{fig:chem_sur} (b)). In contrast, the transverse profile of the orbital truncation rod exhibits two components; a `sharp' one and a `broad' one. Such a line shape is expected when a surface --- in this case the orbital surface --- has a well-defined average position (a flat plane) but local, finite deviations from this average\cite{Andrews85JPhysC}. Thus, without recourse to any model-dependent fitting we can immediately conclude from the data of Fig.\ref{fig:OO-CTR} (a) that there exists  a flat chemical surface and a rougher orbital order surface.

A rougher surface also has the consequence that the intensity along the truncation rod must fall more sharply as a function of $L$ than for a less rough surface. Figure~\ref{fig:OO-CTR} (b) shows the intensity of the
(20$L$) and ($\frac 94 \frac 14 L$) rods as a function of $L$ measured
on two samples (cleaved in He, sample A, and in air, sample B) at 100K, 170K, and 200K. For the orbital rod, the intensity of the sharp component integrated in the transverse direction is shown. Note, the $L$-dependence of the (20$L$) intensity of sample A is slightly different from that of sample B, suggesting a slightly different degree of flatness of the chemical surface for the two samples. On the
other hand, the ($\frac 94 \frac 14 L$) intensity distribution for both
samples is very similar and in both cases, the slope is steeper than that along (20$L$). This is consistent with the result in Fig.~\ref{fig:OO-CTR} (a):
the orbital order surface is rougher than the chemical surface.

\begin{figure}
\includegraphics[width=7cm]{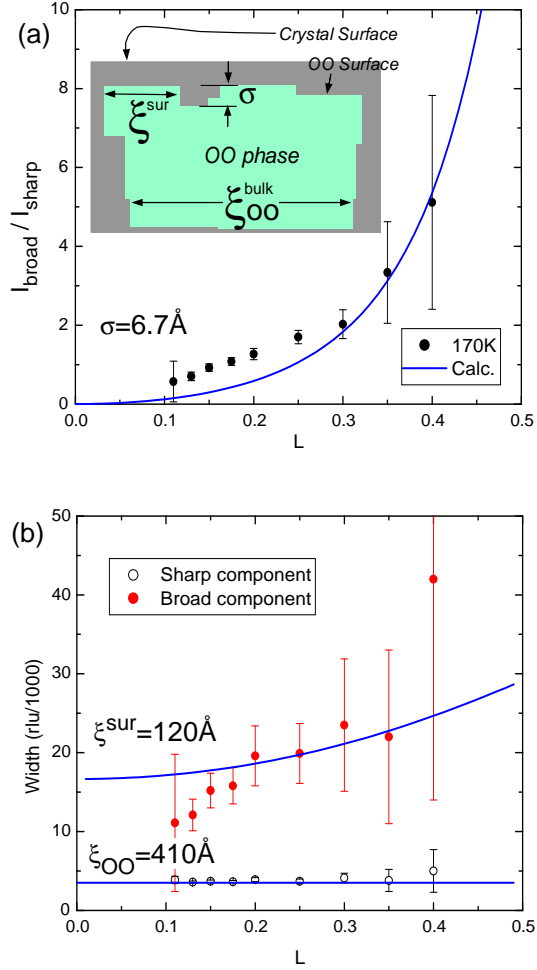}
\caption{(a) The intensity ratio of the broad component observed in the transverse orbital truncation rod scans to the sharp component, as a function of $L$. (inset) Schematic view of the orbital order region around the surface. (b) Peak widths of broad- and sharp- component. Calculated values for the model discussed in the text are shown by solid lines in both figures.}
\label{fig:analysis}
\end{figure}

We next quantify this result by use of a model for a nearly flat surface due to Andrews and Cowley\cite{Andrews85JPhysC}. In this model, the surface is characterized by
two parameters: $\sigma$, the standard deviation of the surface height,
and $\xi^{\scriptsize {\rm sur}}$, the correlation length of the surface (Fig. \ref{fig:analysis}). The truncation rod intensity distribution $I(\vec Q)$ from such a surface is given by:
\begin{eqnarray}
I(\vec Q)&=&I_s(\vec Q) + I_b(\vec Q),\label{eq:I}\\
{\rm with \hspace{1cm}} I_s(\vec Q)&=&\frac{4\pi^2|F(\vec Q)|^2}{q_z^2}\exp(-\sigma^2 q_z^2) \delta(h,k),
   \label{eq:Is}\\
{\rm and \hspace{1cm}} I_b(\vec Q)&=&\pi|F(\vec Q)|^2\frac{(\xi^{\scriptsize {\rm sur}})^2 [1-\exp(-\sigma^2q_z^2)]^2}{\sigma^2q_z^4} 
\nonumber\\&&\times
              \exp\left[\frac{-(\xi^{\scriptsize {\rm sur}})^2[1-\exp(-\sigma^2q_z^2)]}{4\sigma^2q_z^2}(h^2+k^2)\right],\label{eq:Ib}
\end{eqnarray}
where $I_s$ and $I_b$ are the intensities for the sharp component and
the broad component respectively, $F(\vec Q)$ is the structure factor at scattering vector $\vec Q$, and $q_z$ is the $c$*-component of the reduced
wavevector (e.g. this is $2\pi L/c$ for $|L|<0.5$). In order to compare our data to the predictions of this model, we replace the delta-function in equation (\ref{eq:Is}) with a Gaussian, to account for the finite bulk orbital correlation length, and fit the transverse scans to a two-Gaussian lineshape as shown by the red solid curve in Fig.~\ref{fig:OO-CTR}(a). The $L$-dependence of the parameters derived from these fits are shown in Figure \ref{fig:analysis}(a), which plots the ratio of the intensity of the broad component and the sharp component, and in Figure \ref{fig:analysis}(b), which shows the widths of the two components, for the 170~K data set.
The intensity ratio $I_b/I_s$ increases rapidly with $L$. This
ratio depends mainly on $\sigma$, and hence $\sigma$ may be obtained from the plot. The solid line in Fig.~\ref{fig:analysis}(a) shows the expected behaviour for $\sigma=6.7$\AA,\/ which reproduces the data quite well. Note, this same value of $\sigma$ is also consistent with the results shown in Fig.~\ref{fig:OO-CTR} (b); The solid line in this latter figure is the calculated intensity distribution given by taking the product of the measured $I(20L)$ and $\exp (-\sigma^2q_z^2)$, with $\sigma$=6.7\AA\/; here, we assumed the chemical surface is perfectly flat. The in-plane correlation length $\xi^{\scriptsize {\rm sur}}$ is given by the transverse peak width and is shown in Fig.~\ref{fig:analysis} (b) as a function of $L$, together with the $L$-dependence predicted by eq. (\ref{eq:Ib}) for a particular value of $\xi^{\scriptsize {\rm sur}}$. Using the value of $\sigma=6.7$\AA, $\xi^{\scriptsize {\rm sur}}$ is found to be 120\AA. That is, the surface correlation length is much shorter than that of the bulk orbital order $\xi_{OO}^{\scriptsize{\rm bulk}}=410$\AA. This suggests that some aspect of the surface disrupts the orbital order. This may be due to differing surface energetics resulting from the reduced coordination, or perhaps varying Sr concentrations at the site of the cleave.

We next compare the present results to that of previous reports. LEED measurements on the same material \cite{Plummer01ProgSurfSci} showed no superlattice intensity down to 80K, implying little orbital order is established in the topmost layer. Our result is consistent with this study. The surface roughness in the orbital order observed here necessarily requires a small order parameter or a small ordered volume fraction at the surface. The same authors also performed STM measurements on the same system and found unidentified `electronic roughness at
the surface' having an in-plane correlation length of several nanometres
and surface roughness of 6\AA. Comparing to the present results, one notes that the length scale of the roughness observed in the STM study is very
similar to that of the orbital order interface observed in this
study, and it is tempting to surmise that this `electronic roughness' may be related in some way to the rough orbital interface observed here. However, it is difficult to draw any firm conclusions on this point without further study. 

\begin{figure}
\includegraphics[width=7cm]{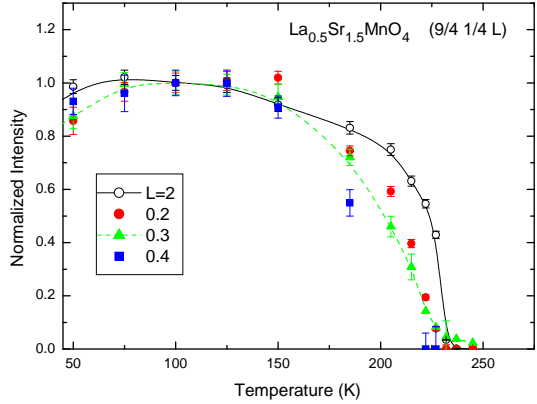} \caption{Temperature dependence of
the ($\frac 94 \frac 14 L$)-sharp component intensity for $L=$2, 0.2,
0.3, and 0.4.}  \label{fig:Tdep}
\end{figure}

Finally, we briefly discuss the temperature dependence of the orbital truncation rod scattering. Figure \ref{fig:Tdep} shows the temperature dependence of the
($\frac 94 \frac 14 L$)-sharp component intensity for $L=$2, 0.2, 0.3,
and 0.4 r.l.u. 
The temperature dependence of the surface-sensitive
OTR scattering intensity ($L$=0.2, 0.3 and 0.4 data) is different from that of the bulk-sensitive
superlattice reflection. 
The bulk OO transition was reported to be `probably second order' with a relatively discontinuous transition\cite{Larochelle05PRB}.
In contrast, the temperature dependence of the OTR scattering intensity is seen to evolve much more continuously through the transition.
 The bulk superlattice reflection intensity is proportional to the volume of the orbitally ordered phase, and to the square of the order parameter, or the Jahn-Teller distortion caused by the orbital ordering. In contrast, the OTR intensity is proportional to the area of the orbital ordered flat surface and to the square of the order parameter. Since the surface roughness of the OO phase at 100K is the same as that at 200~K, then the change in intensity must be attributed to a change in the magnitude of the atomic displacement, i.e., in the orbital order parameter at the surface. Thus, these data suggest that the surface orbital order parameter decays more continuously than the bulk. Such phenomena may be related to surface-induced second order phase transitions that have been observed in other materials\cite{Watson96PRL,Dosch88PRL}. Future work will investigate this possibility in detail.

In conclusion, we have performed a surface x-ray scattering study of a
cleaved, single-layered, half-doped manganite. The chemical surface was
found to be extremely flat with a La/Sr layer termination. Importantly, we have
observed the scattering arising from truncation of the orbital order at a surface for the first time. We find that the orbital surface is rougher than the chemical surface, and that its temperature dependence is different from the bulk orbital order, indicating a surface-induced second order transition. These results are further evidence of the importance of surface effects in determining electronic ground states in strongly correlated systems and provide a direct route to probe such effects quantitatively. Detailed understanding will require theoretical models capable of dealing with such surface effects explicitly.

\section*{Methods}
X-ray diffraction measurements were performed at the beamlines X22C and
X21 at the National Synchrotron Light Source, and at 6ID at the Advanced
Photon Source. The measurements at X22C and 6ID were performed with 6.5 keV
x-rays using a six-circle diffractometer for surface scattering
experiments and those at X21 were performed with 8.8 keV x-rays and a
standard four-circle diffractometer. Most of the experiments using the six-circle diffractometers were conducted with a fixed-$\alpha$ mode in
order to keep the illuminated sample volume constant, where $\alpha$
denotes the incidence angle, i.e., the angle between the crystal surface
and the incident x-ray. The angle $\alpha$ was fixed to 0.6$^\circ$,
which corresponds to a penetration depth of 600 \AA\/, unless
otherwise noted. This angle is higher than the critical angle for total
reflection ($\theta_C=0.4^\circ$). The experiments at X21 were performed
in a $\alpha =\beta$ mode, where $\beta$ denotes the exit angle. Both
modes give very similar results. The temperature was controlled by
closed-cycle He refrigerators. The sample was mounted either in
thermally-insulating vacuum or in a He exchange gas.
The curve fitting for CTR profile was made with the ROD program\cite{Vlieg00JApplCryst} with Mo for La/Sr site because the form factor of Mo approximates that of the La$_{0.25}$Sr$_{0.75}$ mixture well.

\section*{Acknowledgement}
We thank Dr. A. Checco for helping with AFM measurements. Y.W. wishes to acknowledge the Yamada Science Foundation, support for long-term visit. US DOE, Basic Energy Sciences supported work at Brookhaven under contract No.~DE-AC02-98CH10886, at Argonne under contract No.~DE-AC02-06CH11357, and at the MUCAT Sector at the Adcanced Photon Source and Ames Laboratory under contract No.~DE-AC02-07CH11358.

\end{document}